\documentclass[a4paper]{article}

\usepackage{INTERSPEECH2020}
\usepackage{multirow}

\title{Investigation of Phase Distortion on Perceived Speech Quality for Hearing-impaired Listeners}
\name{Zhuohuang Zhang$^{1,2}$, Donald S. Williamson$^2$, Yi Shen$^1$}
\address{
  $^1$Department of Speech, Language and Hearing Sciences, Indiana University, USA\\
  $^2$Department of Computer Science, Indiana University, USA}
\email{zhuozhan@iu.edu, $\{$williads, shen2$\}$@indiana.edu}

\begin{document}

\maketitle
\begin{abstract}
  Phase serves as a critical component of speech that influences the quality and intelligibility. Current speech enhancement algorithms are beginning to address phase distortions, but the algorithms focus on normal-hearing (NH) listeners. It is not clear whether phase enhancement is beneficial for hearing-impaired (HI) listeners. We investigated the influence of phase distortion on speech quality through a listening study, in which NH and HI listeners provided speech-quality ratings using the MUSHRA procedure. In one set of conditions, the speech was mixed with babble noise at 4 different signal-to-noise ratios (SNRs) from -5 to 10 dB. In another set of conditions, the SNR was fixed at 10 dB and the noisy speech was presented in a simulated reverberant room with T60s ranging from 100 to 1000 ms. The speech level was kept at 65 dB SPL for NH listeners and amplification was applied for HI listeners to ensure audibility. Ideal ratio masking (IRM) was used to simulate speech enhancement. Two objective metrics (i.e., PESQ and HASQI) were utilized to compare subjective and objective ratings. Results indicate that phase distortion has a negative impact on perceived quality for both groups and PESQ is more closely correlated with human ratings.
\end{abstract}
\noindent\textbf{Index Terms}: phase distortion, speech quality perception, objective metric, human listening study

\section{Introduction}

Phase is a critical component of a speech signal, and it makes important contributions to speech intelligibility and perceived quality. When analyzing a speech signal in the time-frequency (T-F) domain by performing short-time Fourier transform (STFT) analysis, each time-by-frequency location in the resulting spectrogram contains not only magnitude but also phase information. At each time frame, the phase spectrum represents how the various frequency components in the speech signal are temporally aligned. At each frequency, the progression of phase across consecutive time frames represents the temporal fine structure (TFS) of the speech signal, which is important to the perception of talker gender, voicing, and intonation \cite{lorenzi2007role}. Replacing or distorting the phase information when reconstructing speech from the spectrogram would lead to degraded speech intelligibility \cite{xu2017distorting}.

Many of the existing speech enhancement systems only operate based on the magnitude spectrogram and keep the noisy phase unchanged when converting the enhanced speech to the time domain \cite{wang2014training, weninger2014discriminatively, soni2018time}. Recently, a number of studies have shown that better phase estimation of the original speech improves both subjective and objective speech quality \cite{paliwal2011importance, williamson2015complex}. However, these studies addressing the importance of phase in speech enhancement have only focused on normal-hearing (NH) listeners. Approximately one third of older adults above the age of 65 years in the United States suffer from hearing loss \cite{shargorodsky2010change}. Modern digital hearing aids, besides amplifying the acoustic signals, also consist of built-in speech enhancement algorithms to remove unwanted background noise that corrupts the speech \cite{yang2005spectral, zhang2019objective}. This enhancement is performed before amplification. Before the phase-preserving speech enhancement algorithms can be implemented into hearing aids, it is necessary to evaluate whether hearing-impaired (HI) individuals would actually benefit from them.



It is known that HI listeners have poorer sensitivity to the TFS \cite{buss2004temporal, moore2006frequency} and benefit less from TFS cues for speech understanding \cite{lorenzi2006speech, hopkins2008effects}.  Therefore, it may be expected that  preserving the phase information in speech enhancement may not lead to the same degree of benefit for HI listeners compared to NH listeners. In order to optimize speech enhancement algorithms for the HI population, it is crucial to quantify their sensitivity to phase distortions, especially if the phase distortions remained in the enhanced speech following phase-insensitive enhancement. If HI listeners consistently rate the phase-distorted speech as having lower quality, even after phase-insensitive enhancement, then it would suggest the potential benefit for a phase-sensitive system. Furthermore, objective speech-quality metrics developed for NH listeners may not directly generalize to HI listeners \cite{kates2014hearing}. Therefore, it is not clear whether existing speech-quality metrics could be adequately used to capture the effect of phase distortion for HI listeners.

To address these open questions, we collected human quality ratings on noisy speech signals with different degrees of phase distortion (or equivalently distortion to the TFS) for both NH and HI listeners using the MUltiple Stimuli with Hidden Reference and Anchor (MUSHRA) procedure \cite{itu2014recommendation}. The quality ratings were repeated on speech signals with and without processing from a common phase-insensitive algorithm based on an ideal ratio mask (IRM) in the T-F domain \cite{wang2014training}. This allowed us to investigate whether the perceived quality by NH and HI listeners would be adversely affected if phase distortion remained in the enhanced speech following traditional, magnitude-based enhancement. Comparing the ratings from these two conditions also reveals the expected benefits from a phase-insensitive speech enhancement system. To assess the agreement between subjective and objective speech quality, the subjective quality ratings were compared against two objective metrics, namely perceptual evaluation of speech quality (PESQ) \cite{rix2001perceptual} and hearing-aid speech quality index (HASQI) \cite{kates2014hearing}.


In the following, the implementation of phase distortion and the procedure for the subjective listening test are described in Section 2. In Section 3, the results obtained from the listening test and their correlations to the objective metrics are presented. Finally, conclusions are drawn in Section 4.

\section{Methods}
\subsection{Phase distortion}
Phase distortion was artificially applied to the speech materials by introducing random perturbations to the phase spectrogram using the following steps. First, both the magnitude spectrogram [$|s(t,f)|$] and the phase spectrogram [$\angle s(t,f)$] in the T-F domain were extracted. A window size of 25 ms with a step size of 10 ms were used during STFT analysis. Second, four degrees (i.e., 25\%, 50\%, 75\%, and 100\%) of phase distortion were applied to the phase spectrogram according to
\begin{equation}
\angle s(t,f)_{\text{distorted}} = \angle s(t,f) + \alpha \cdot \phi(t,f),
\end{equation}
where $\angle s(t,f)_{\text{distorted}}$ denotes the distorted phase in the T-F domain; $\alpha$ denotes the amount of phase distortion ranging from 25\% to 100\%; and $\phi(t,f)$ represents random phase perturbations drawn from a uniform distribution between 0 and $2\pi$, independently for each T-F location. Similar distortion amounts were used in earlier studies involving NH listeners~\cite{mathesmiller1947, CraigJeffress1962, plompSteeneken1969}. These amounts also reflect potential errors due to inaccurate phase enhancement. Finally, the phase-distorted speech was resynthesized with the inverse STFT that combines the original magnitude spectrogram and the distorted phase spectrogram. 


\subsection{Subjects}
A total of 18 participants were recruited, including 10 NH listeners (4 males, 6 females, recruited from the undergraduate population at Indiana University) and 8 HI listeners (3 males, 5 females, average age: 68 ($SD = 5.53$)). All participants were native speakers of American English. Audiometric evaluations were performed on all NH listeners, including otoscopy and air-conduction pure-tone audiometry. All NH listeners had audiometric thresholds below 20 dB HL from 250 to 6000 Hz. For the HI listeners, the hearing evaluation additionally included bone-conduction audiometry, tympanometry, and hearing-related case history. All HI listeners had at least mild symmetric hearing loss of a sensorineural origin. The average audiometric thresholds for the two groups are listed in Table \ref{table1}. The current study was conducted following the Declaration of Helsinki and approved by the Institutional Review Board (IRB) at Indiana University. Informed consents were obtained from all participants before the data collection.

\subsection{Stimuli}
Speech utterances from the IEEE corpus \cite{rothauser1969ieee} produced by a female talker were adopted for the listening test. To ensure speech audibility for HI listeners, a standard hearing-aid prescription formula (i.e., NAL-R) \cite{byrne1986national} was used to amplify the speech stimuli. This formula prescribes linear gains for various frequency regions according to the degrees of hearing loss in these regions. The speech level was calibrated to 65 dB SPL before amplification. All signals were resampled to 16 kHz before further processing.

\begin{table}[ht]
\caption{Average auditory thresholds of participants from NH and HI groups with the standard deviations in parentheses.}
\vspace{-2mm}
\label{table1}
\centering
\begin{tabular}{l|cccccc}
\multicolumn{7}{c}{\textbf{Average Auditory Thresholds (dB HL)}}  \\ \hline
 & \multicolumn{6}{c}{\textbf{Frequency (kHz)}} \\ 
 & .25 & .5 & 1 & 2 & 4 & 6 \\ \hline
 \textbf{NH} & 10.0 & 8.5  & 6.0  & 5.0 & 5.0 & 11.0 \\
            & (4.7) &  (5.3) &  (3.9) & (5.3) & (6.7) & (6.2) \\
 \textbf{HI} & 23.1 & 20.6 & 26.9 & 36.3 & 44.4 & 46.3 \\
             & (8.0) &  (9.0) & (14.1) & (18.5) & (19.7) & (16.6) 
\end{tabular}
\vspace{-4mm}
\end{table}

There were four test conditions in the listening test. In the Noisy condition, the speech stimuli was presented with a simultaneous 10-talker babble from the AzBio database \cite{spahr2012development} at 4 different signal-to-noise ratios (SNRs), from -5 dB to 10 dB with a 5 dB step. In the Noisy-Enhanced condition, the stimuli were the same as those in the Noisy condition except that they were further masked by the IRM \cite{wang2014training} before presentation. In the Reverberant condition, the SNR between the speech and noise was fixed at 10 dB and the stimuli were presented with simulated reverberation~\cite{allenBerkley1979}. The reverberation algorithm simulated a room with a dimension of 4m$\times$4m$\times$3m (length$\times$width$\times$height), the sound source was located at (2m, 3.5m, 2m), and the listener was located at (2m, 1.5m, 2m). The sound velocity was assumed to be 340 m/s. The reverberation times (T60) were 100, 200, 500, and 1000 ms. In the Reverbrant-Enhanced condition, the stimuli were the same as those in the Reverbrant condition except that they were further masked by the IRM before presentation. Note that the IRM in this condition was applied only on the noise without removing reverberation, which resembles a system that is not trained on reverberant data. The noisy and reverberant conditions (with and without enhancement) were chosen since they follow but extend a similar approach that studied the importance of phase for NH listeners \cite{paliwal2011importance}.
 
Since most HI listeners have high-frequency hearing loss, all stimuli were low-pass filtered at 4 kHz. This ensures that the perceived speech quality is not dominated by hearing loss at high-frequency bands. All stimuli were presented monotonically in the participants' better ear based on the hearing screening results. A 24-bit soundcard (Microbook II, Mark of the Unicorn, Inc.) and a pair of headphones (HD280 Pro, Sennheiser electronic GmbH and Co. KG) were used. The participants were seated in a sound-attenuating booth during the study.

\subsection{Procedure}
Listeners provided subjective ratings on the stimuli following the MUSHRA procedure, recommended in ITU-R BS.1534 \cite{itu2014recommendation}. During the experiment, a graphic user interface (GUI) was shown on a computer screen in front of the listener. The user interface contained a ``Reference" button that corresponded to a reference stimulus, which was the original clean speech low-pass filtered at 4 kHz. The six additional buttons representing the six test stimuli, which were six versions of the same sentence stimulus as the reference. One of these button corresponded to the reference stimulus; one button corresponded to a hidden anchor stimulus, which was the original clean speech low-pass filtered at 2 kHz; the remaining four buttons corresponded to the phase-distorted speech with four degrees of phase distortion. The correspondence between the buttons and the stimuli was randomized from trial to trial. On each trial, the listener clicked on each of the buttons to hear the corresponding speech stimulus and rate the quality of the stimulus on a scale from 1 to 100 using a slider next to the button. The listener was instructed that the quality of the reference stimulus corresponded to a rating of ``100". The listener was able to play the reference and test stimuli more than once. 

The experiment was completed in two 2-hour sessions. For half of the listeners in each of the two listener groups, the Noisy and Noisy-Enhanced conditions were tested in the first session while the Reverberant and Reverberant-Enhanced conditions were tested in the second session. For the other half of the listeners, the test sequence for the two sessions was reversed. At the beginning of each session, eight practice trials were run to familiarize the listener with the stimuli and the GUI. If the Noisy and Noisy-Enhanced conditions were tested in the session, the practice trials included stimuli at the four different SNRs, with and without the IRM-based enhancement. After the practice trials, the two experimental conditions were tested using two blocks of 40 trials. The order in which the two conditions were tested was counterbalanced across listeners. Within each block, 10 trials were run at each of the SNRs, in random order. If the Reverberant and Reverberant-Enhanced conditions were tested in the session, the practice trials included stimuli at the four different values of T60, with and without the IRM-based enhancement. Following the practice trials, the two experimental conditions were tested in blocks, with each block containing 10 trials at each of the T60 values in random order. No sentence was repeated in more than one trial, leading to a total of 176 unique sentences used in the current experiment. Due to the limited availability, one HI listener did not finish the Reverberant and Reverberant-Enhanced conditions and another HI listener did not finish the Noisy and Noisy-Enhanced conditions, resulting in seven listeners in the HI group for each condition. For data collected from each session, a mixed-effect analysis of variance (ANOVA) was conducted to identify any significant effects of listener group, speech enhancement, SNR/T60, phase distortion and any significant interactions among them.

Two objective quality metrics, PESQ and HASQI, were adopted to further investigate the correlations between objective measures and actual human ratings, especially on speech signals with distorted phase under noisy and reverberant conditions. PESQ is a widely adopted metric for speech quality assessment that gives outputs ranging from -0.5 to 4.5, while HASQI is a more recently proposed speech quality metric that includes a physiologically inspired model of human auditory system with predicted scores ranging from 0 to 1. The inclusion of this model allows HASQI to predict the perceived quality by both NH and HI listeners. The same stimuli (with NAL-R linear amplification and low-pass filtering) presented to each subject were given as inputs to both evaluation metrics. 

\section{Results and discussion}

\begin{figure}[htb!]
  \centering
  \includegraphics[width=\linewidth]{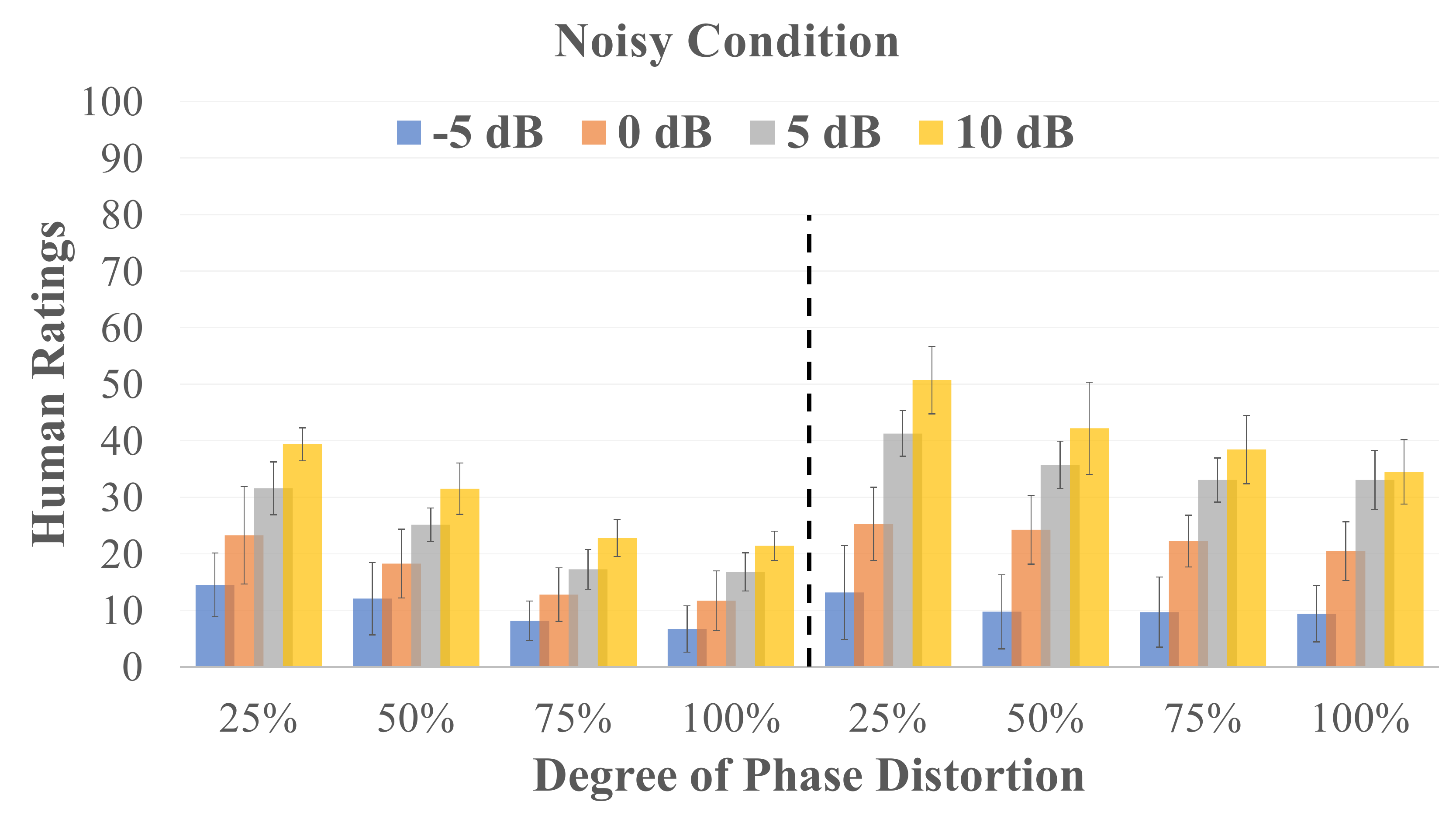}
  \caption{Human ratings under Noisy condition, NH listeners are represented in the \textbf{left} block and the HI listeners are shown in the \textbf{right} block.}
  \label{fig:noisy}
  \vspace{-3.5mm}
\end{figure}

\begin{figure}[htb!]
  \centering
  \includegraphics[width=\linewidth]{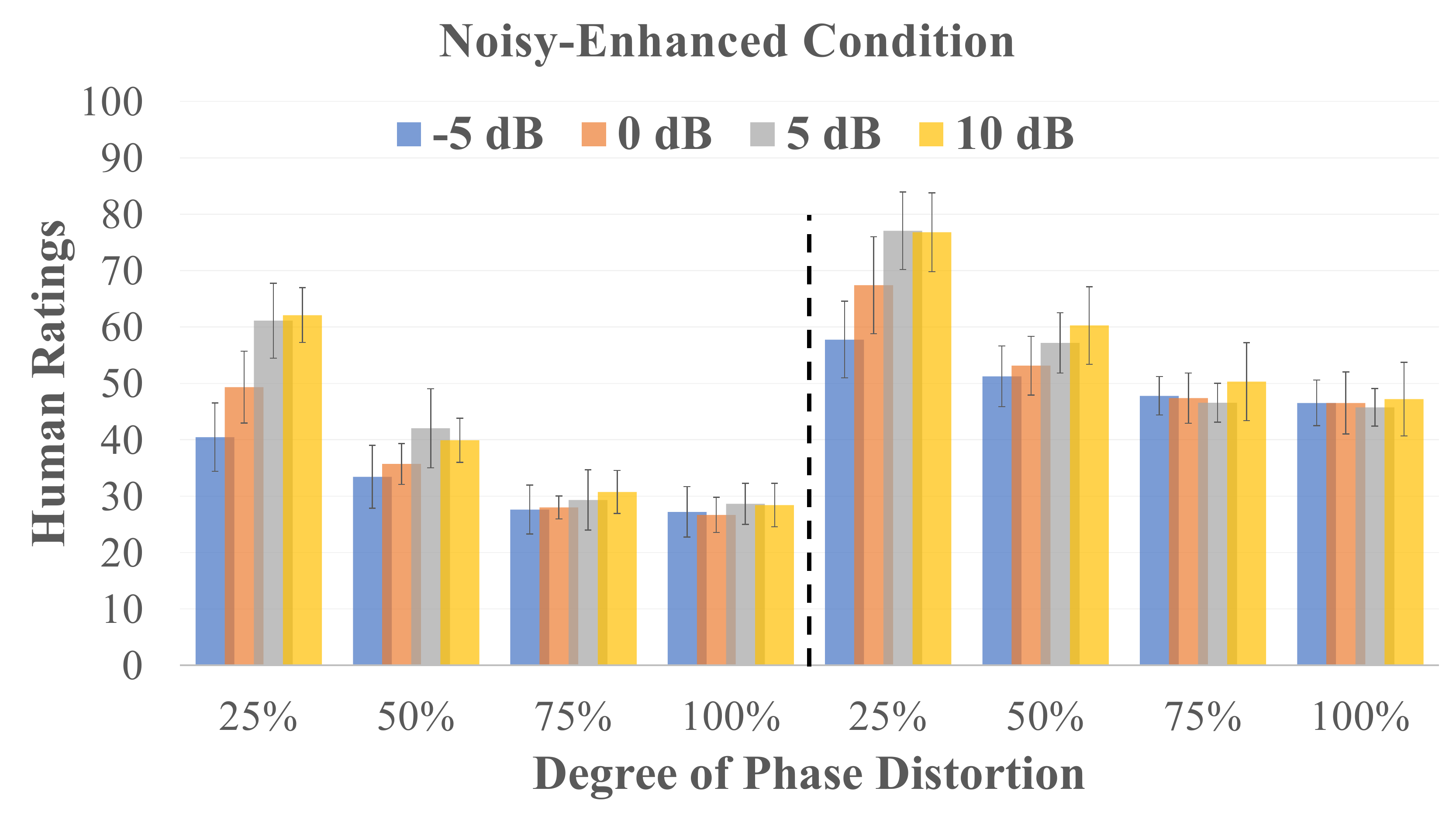}
  \caption{Human ratings under Noisy-Enhanced condition, NH listeners are represented in the \textbf{left} block and the HI listeners are shown in the \textbf{right} block.}
  \label{fig:noisy-enhanced}
  \vspace{-5mm}
\end{figure}

Figs.~\ref{fig:noisy} and \ref{fig:noisy-enhanced} show the subjective ratings for the four different degrees of phase distortion and the four SNRs in the Noisy and Noisy-Enhanced conditions. The error bars indicate $\pm$ one standard deviation. A mixed-design ANOVA shows significant main effects of listener group [$F(1,15)=6.35, p=.024$], enhancement [$F(1,15)=68.14, p<.001$], SNR [$F(1.7, 25.1)=90.67, p<.001$, Greenhouse-Geisser corrected], and phase distortion [$F(1.1, 16.6)=77.33, p<.001$, Greenhouse-Geisser corrected]. There are significant interactions between listener group and SNR [$F(1.7, 25.1)=3.21, p=.032$, Greenhouse-Geisser corrected], between enhancement and SNR [$F(1.5, 22.7)=24.54, p<.001$, Greenhouse-Geisser corrected], between enhancement and phase distortion [$F(1.3, 19.9)=47.86, p<.001$, Greenhouse-Geisser corrected], and between SNR and phase distortion [$F(9,135)=25.08, p<.001$], as well as significant three-way interactions among listener group, enhancement, and SNR [$F(1.5, 22.65)=5.59, p=.016$, Greenhouse-Geisser corrected] and among enhancement, SNR, and phase distortion [$F(3.8, 56.9)=4.85, p<.001$, Greenhouse-Geisser corrected]. 


For the Noisy condition (Fig.~\ref{fig:noisy}), higher SNRs lead to higher quality ratings and greater degrees of phase distortions lead to lower ratings for both listener groups. The effects of SNR and phase distortion also interact with each other, with stronger effects of phase distortion observed at higher SNRs. When the noise level is high (i.e., at low SNRs), the phase distortion may be masked by the noise and become less noticeable to the listeners. The HI listeners tend to give higher ratings than the NH listeners, suggesting that they have higher tolerance for phase distortion and background noise. 

For the Noisy-Enhanced condition (Fig.~\ref{fig:noisy-enhanced}), the quality ratings are generally higher than those in the Noisy condition, indicating that the IRM-based enhancement improved perceived quality. Contrary to the strong effect of SNR in the Noisy condition, the effect of SNR is not reliably observed in the Noisy-Enhanced condition across all phase distortions. This suggests that following phase-insensitive enhancement the contribution from the background noise to the quality ratings is much reduced. Instead the quality ratings become dominated by phase distortion especially when the distortion amount is above 25\% for the enhanced speech. For both listener groups, the quality rating decreases as the degree of phase distortion increases. In particular, the enhancement algorithm allows both the NH and HI listeners to better differentiate various degrees of phase distortion at low SNRs. Therefore, a speech enhancement algorithm that is capable of reducing phase distortions would likely lead to benefits in perceived speech quality for all listeners (with or without hearing loss), at both high and low SNRs.

\begin{figure}[htb!]
  \centering
  \includegraphics[width=\linewidth]{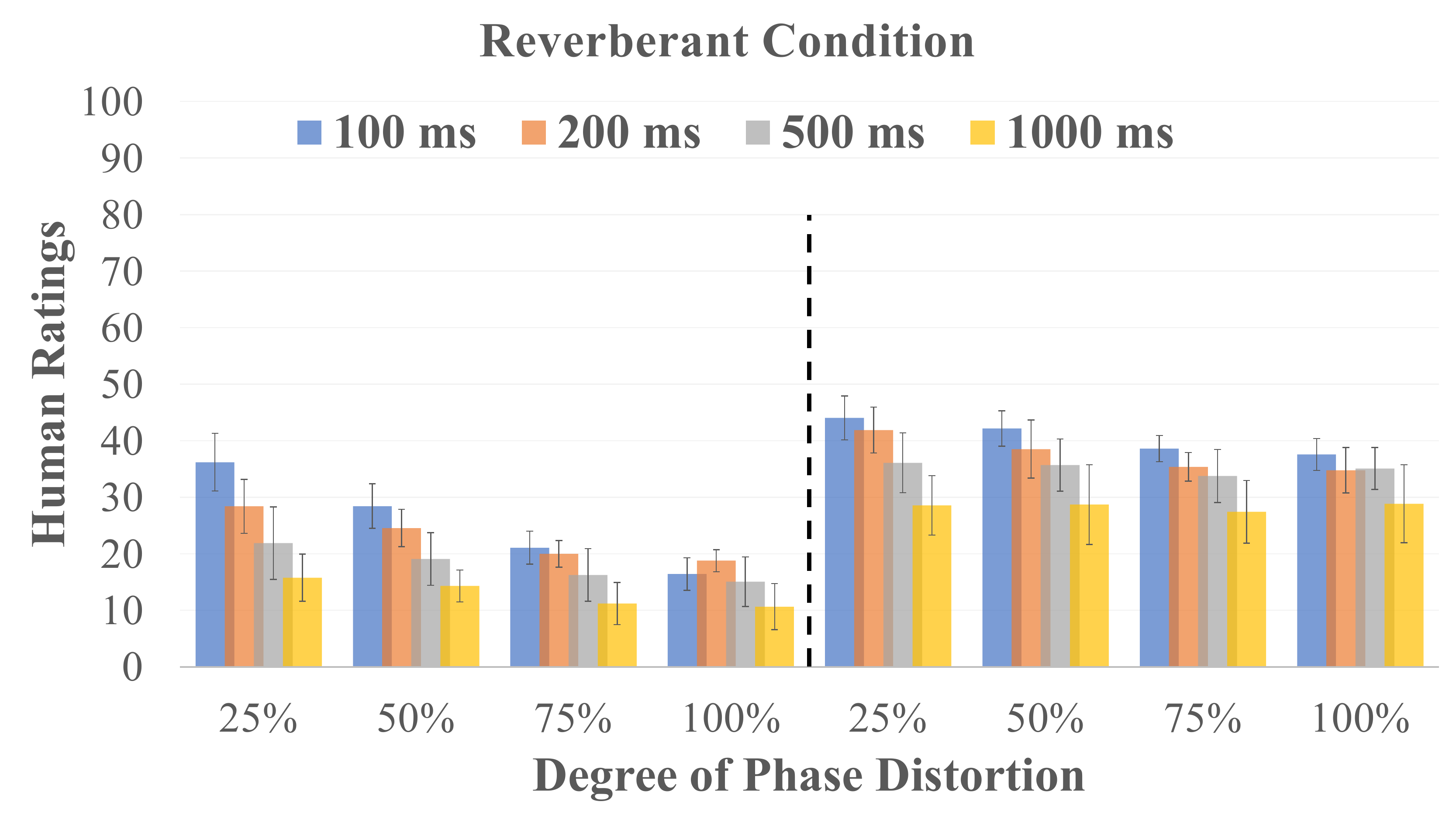}
  \caption{Human ratings under Reverberant condition, NH listeners are represented in the \textbf{left} block and the HI listeners are shown in the \textbf{right} block.}
  \label{fig:reverb}
  \vspace{-3.5mm}
\end{figure}
\begin{figure}[htb!]
  \centering
  \includegraphics[width=\linewidth]{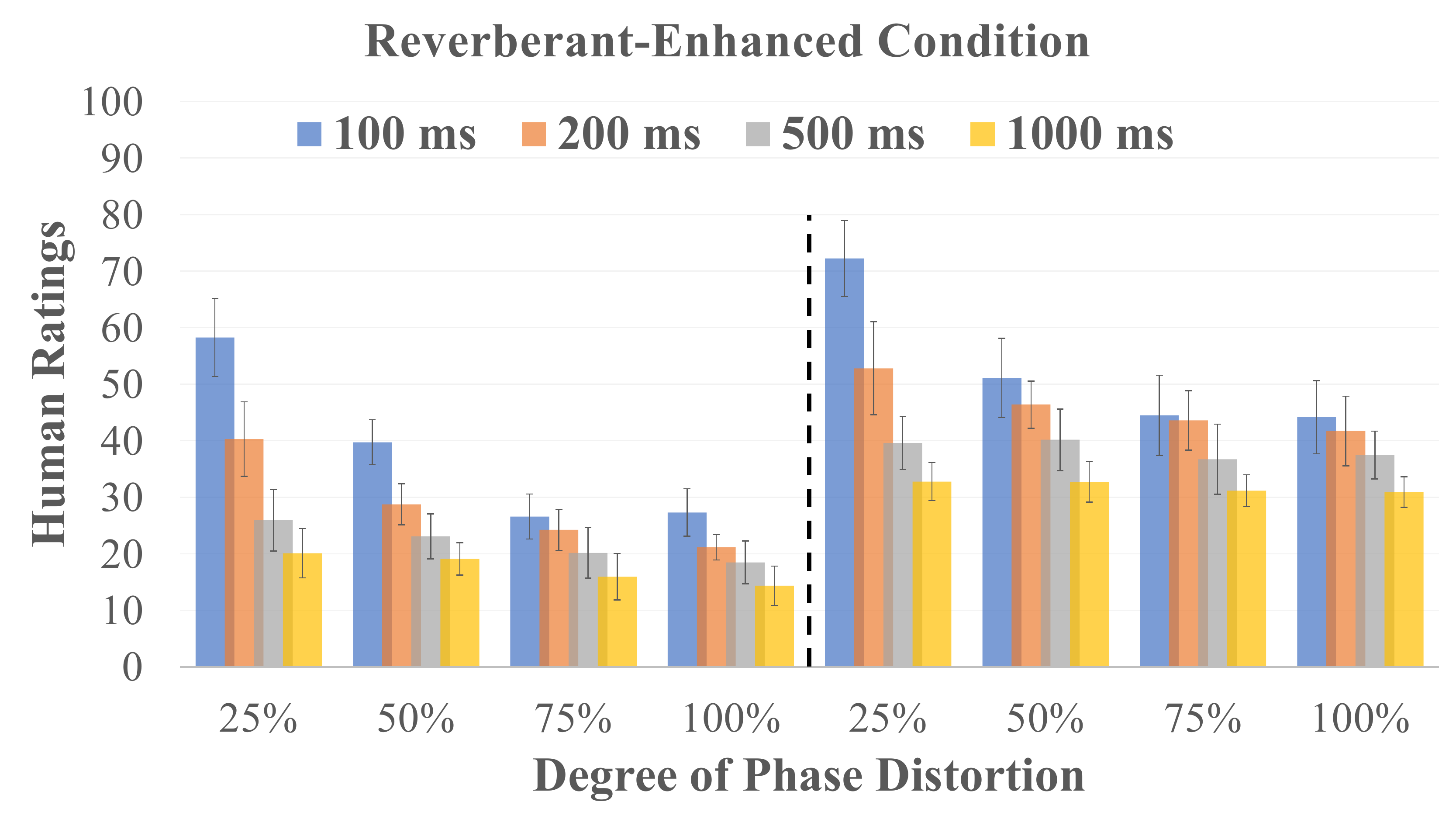}
  \caption{Human ratings under Reverberant-Enhanced condition, NH listeners are represented in the \textbf{left} block and the HI listeners are shown in the \textbf{right} block.}
  \label{fig:reverb-enhanced}
  \vspace{-5mm}
\end{figure}

Figs.~\ref{fig:reverb} and \ref{fig:reverb-enhanced} show the subjective ratings for the four different degrees of phase distortion and the four T60 values in the Reverberant and Reverberant-Enhanced conditions. The error bars indicate $\pm$ one standard deviation. A mixed-design ANOVA shows significant main effects of listener group [$F(1,15)=13.51, p=.002$], enhancement [$F(1,15)=8.92, p=.009$], reverberation [$F(1.9, 27.8)=48.60, p<.001$, Greenhouse-Geisser corrected], and phase distortion [$F(1.2, 18.1)=51.74, p<.001$, Greenhouse-Geisser corrected]. There are significant interactions between enhancement and reverberation [$F(3, 45)=12.12, p<.001$], between enhancement and phase distortion [$F(1.3, 19.0)=9.87, p<.001$, Greenhouse-Geisser corrected], between reverberation and phase distortion [$F(2.8,42.7)=30.63, p<.001$, Greenhouse-Geisser corrected], as well as significant three-way interactions among enhancement, reverberation, and phase distortion [$F(3.6, 53.7)=11.59, p<.001$, Greenhouse-Geisser corrected].



For the Reverberant condition (Fig.~\ref{fig:reverb}), shorter reverberation times lead to higher quality ratings and greater degrees of phase distortions lead to lower ratings for both listener groups. The effects of reverberation time (T60) and phase distortion also interact with each other, with stronger effects of phase distortion observed for shorter reverberation times. It is possible that long reverberation times (e.g., 1000 ms) could mask the phase distortion applied to the speech and make it less noticeable. 


For the Reverberant-Enhanced condition (Fig.~\ref{fig:reverb-enhanced}), the quality ratings are generally higher than those in the Reverberant condition, indicating that the IRM-based enhancement improved perceived speech quality. Following speech enhancement, the effect of T60 is still present for all degrees of phase distortion. This suggests that the IRM-based enhancement, when trained using non-reverberant noise and speech, is insufficient in removing the adverse effect of reverberation on speech quality. Moreover, following speech enhancement the effect of phase distortion becomes stronger for short reverberation times (100 and 200 ms). On the other hand, because of the remaining reverberation in the enhanced speech, the effect of phase distortion is absent for long reverberation times (1000 ms). The dependencies of the quality rating on phase distortion are similar between the NH and HI groups, suggesting that both NH and HI listeners could benefit from speech enhancement techniques that restore the phase spectrogram of the original clean speech.


The subjective speech quality ratings obtained from the listening experiment were compared to two objective metrics, PESQ and HASQI. The correlations between the subjective and objective speech quality are listed in Table.~\ref{tab:correlation}. PESQ yields the highest Pearson correlation coefficients for both NH and HI groups across all conditions. HASQI achieves similar correlation coefficients for the Noisy and Reverberant conditions. This is contrary to the reports in \cite{kressner2011robustness,wirtzfeld2017predicting}, where HASQI yields better correlation than PESQ for enhanced speech. We infer that the frequency range for evaluation might be one of the factors that contribute to HASQI's failure, since PESQ is designed for measuring signal quality transmitted through a narrow band (3.1 kHz) \cite{rix2001perceptual}, while HASQI is designed for a wider frequency band (12 kHz) \cite{kates2014hearing}. The low-pass filtering applied in our experiment makes PESQ more fitted to this scenario and previous studies adopted wide-band stimuli that is more suitable for HASQI.

\begin{table}[]

\caption{Pearson correlations between subjective and objective ratings at different conditions for NH and HI groups. Highest correlations are marked in \textbf{bold}. `Orig.' indicates the original conditions before enhancement. `Reverb.' stands for reverberant conditions.}
\vspace{-2mm}
\label{tab:correlation}
\centering
\begin{tabular}{lccccc}
\multicolumn{6}{c}{\textbf{Pearson Correlation Coefficients}}                                                                        \\ \hline
                        & \multicolumn{1}{l|}{}   & \multicolumn{2}{c|}{PESQ}                            & \multicolumn{2}{c}{HASQI} \\
                        & \multicolumn{1}{l|}{}   & Orig.          & \multicolumn{1}{c|}{Enhanced}       & Orig.      & Enhanced     \\ \hline
\multirow{2}{*}{Noisy}  & \multicolumn{1}{c|}{NH} & \textbf{0.984} & \multicolumn{1}{c|}{\textbf{0.975}} & 0.956      & 0.755        \\
                        & \multicolumn{1}{c|}{HI} & \textbf{0.991} & \multicolumn{1}{c|}{\textbf{0.971}} & 0.976      & 0.871        \\ \hline
\multirow{2}{*}{Reverb.} & \multicolumn{1}{c|}{NH} & \textbf{0.993} & \multicolumn{1}{c|}{\textbf{0.992}} & 0.955      & 0.883        \\
                        & \multicolumn{1}{c|}{HI} & \textbf{0.990} & \multicolumn{1}{c|}{\textbf{0.986}} & 0.974      & 0.940       
\end{tabular}
\vspace{-5mm}
\end{table}

\vspace{-2mm}
\section{Conclusions}

We investigated the influence of phase distortion on the perceived speech quality in NH and HI listeners. Hearing-impaired listeners tend to provide higher ratings for the same speech stimulus, corrupted by either background noise or reverberation, than NH listeners. However, the quality rating depends on the degree of phase distortion in a similar way. Following phase-insensitive enhancement, HI and NH listeners can differentiate the degree of phase distortion that remained in the enhanced speech, indicating potential benefits from phase-sensitive enhancement techniques. We surmise that these HI listeners may notice phase distortions because (1) they have good TFS sensitivity, or (2) TFS and phase cues are weighed higher for quality tasks as compared to recognition tasks. Future efforts will address these points. Between two objective speech-quality metrics, PESQ provides closer correlations to the subjective ratings than HASQI, especially for the enhanced speech.


\vspace{-2mm}
\section{Acknowledgements}

This work was supported by the Indiana University Vice Provost for Research through the Faculty Research Support Program (co-PIs: D. S. Williamson and Y. Shen) and by an NIH grant R01DC017988 (PI: Y. Shen). We thank James Kates for providing the code for HASQI. We also thank Jillian E. Bassett, Bailey E. Henderlong, Yi Liu and Annie L.K. Main for their assistance in data collection.

\bibliographystyle{IEEEtran}
\bibliography{mybib}

\end{document}